\documentstyle[aps]{revtex}
\newcommand{\be}{\begin{equation}}
\newcommand{\ee}{\end{equation}}
\newcommand{\bea}{\begin{eqnarray}}
\newcommand{\eea}{\end{eqnarray}}

\newcommand{\pr}{^{\prime}}
\begin{document}
\draft
\baselineskip 16pt
\title{Critical Enhancement of the In-medium Nucleon-Nucleon Cross Section
at low Temperatures}
\author{T. Alm and G. R\"opke}
\address{Arbeitsgruppe der Max-Planck-Gesellschaft
``Theoretische Vielteilchenphysik ``
an der Universit\"at Rostock,\\ Universit\"atsplatz 1, 18051
Rostock, Germany}
\author{M. Schmidt}
\address{Institut f\"ur Ostseeforschung, Seestr. 15, 18119
Rostock-Warnem\"unde, Germany}

\maketitle
\begin{abstract}
\begin{sloppypar}
The in-medium nucleon-nucleon cross section is calculated
starting from the thermodynamic T-matrix
at finite temperatures.
The corresponding Bethe-Salpeter-equation is solved using a
separable representation of the Paris
nucleon-nucleon-potential.
The energy-dependent in-medium N-N cross section
at a given density shows a strong temperature
dependence.
Especially at low temperatures and low total momenta, the
in-medium cross section is strongly modified by in-medium
effects.
In particular, with decreasing temperature an enhancement near
the Fermi energy is observed.
This enhancement can be discussed as a precursor of the
superfluid phase transition in nuclear matter.
\vspace{5mm}

\noindent
PACS numbers: 21.30.+y,21.65.+f,25.70.-z
\vspace{1cm}\end{sloppypar}
\end{abstract}

%\newpage
%%%%%%%%%%%%%%%%%%%%%%%%%%%%%%%%%%%%%%%%%%%%%%%%%%%%%%%%%%%%%%%%%%%%
\section{Introduction}
\label{sec:Int}
%%%%%%%%%%%%%%%%%%%%%%%%%%%%%%%%%%%%%%%%%%%%%%%%%%%%%%%%%%%%%%%%%%%%
Heavy-ion reactions at intermediate energies provide information
about the equation of state of dense nuclear matter in a broad
density and temperature range.
Moreover, they can give hints of the process of
equilibration of the excited  nuclear matter.
The time development of the hot source produced in the course of
a heavy-ion reaction can be simulated in BUU calculations
\protect\cite{Cassing,Ogilvie,Xu}.
An important input for such BUU calculations is the
nucleon-nucleon (N-N) cross
section.
The N-N cross sections entering the collisional integral of the
BUU equation describe the collision of two nucleons within the
hot source, which is formed by the other nucleons produced in the reaction.
Consequently, the use of the free N-N cross section is not
justified.
Instead, the in-medium cross section, that takes into
account the modification of the two-particle scattering process
within a dense medium, has to be used in such simulations \cite{Klakow}.

First results for the in-medium modifications in connection with
the mean free path of a nucleon in hot dense matter
 were obtained by Cugnon et al.
\cite{Cugnon} and by Schmidt et al. \cite{Schmidt}.
The in-medium N-N cross section at zero temperature has been
calculated by Faessler et al. \cite{Faessler,Faessler1}
in the framework of Brueckner theory based on the Reid potential.
They found strong modifications of the free N-N cross section
with increasing density. In particular, their calculations showed
a non-monotonous behaviour of the cross section with the density.
Ter Haar et al. \cite{Mal} determined the in-medium cross section
in the frame-work of the relativistic Dirac-Brueckner approach at
$T=0$, also showing strong deviations from the free cross
section.
Using the Bonn N-N-potential within the Dirac Brueckner approach
Li et al. \cite{Li} calculated the
in-medium N-N cross section in nuclear matter at zero temperature.
In  refs. \cite{Mal}, \cite{Li} a substantial reduction of
the in-medium cross section, particularly for low energies, was found.
%%%%%%%%%%%%%%%%%%%%%%%%%%%%%%%%%%%%%%%%%%%%%%%%%%%%%%%%%%%%%%%%%%%%%%

Within this paper, we use a thermodynamic Green's function
approach with non-relativistic propagators to determine the
in-medium N-N cross section at finite temperatures.
The calculations have been performed in
the ladder approximation for the
thermodynamic T-matrix (including Pauli blocking and self energy
contributions) that is solved using a separable representation of
the Paris N-N-interaction.
These methods have already been applied to the calculation of
the thermodynamic properties of dense nuclear matter
in ref. \cite{Schmidt}.
We discuss the dependence of the in-medium N-N cross section
on
the thermodynamic parameters, i.e. density and temperature, and
on the total momentum of the pair.
A systematic calculation of the in-medium N-N cross section in a
broad parameter range will be given in  a separate publication
\cite{AlmB} where also the consequences for BUU simulations of
heavy-ion reactions will be discussed.
Here we will concentrate on the behaviour of the in-medium
cross section at low temperatures ($T<10$ MeV).

It is wellknown,
that in this temperature range a phase transition of nuclear matter
into a superfluid state is expected from BCS-theory (see
e.g. ref.  \cite{Baldo} and references therein).
The ladder approximation for the thermodynamical T-matrix gives
on one hand side the in-medium cross sections at finite
temperatures, on the other hand side it permits the
determination of the critical temperature for the superfluid
phase transition (Thouless criterion \cite{Thouless}).
We consider the relation between the behaviour of the
in-medium cross section at low temperatures and the onset of a superfluid
state in nuclear matter in detail.
In particular, we will show that the temperature at
which the in-medium cross section exhibits a  divergency
(for a given density and zero total momentum of the pair) is identical
with the critical temperature found from the Thouless criterion
for the onset of superfluidity in symmetric nuclear matter \cite{Alm1}.

%%%%%%%%%%%%%%%%%%%%%%%%%%%%%%%%%%%%%%%%%%%%%%%%%%%%%%%%%%%%%%%%%%%%%%%%%%%

\section{The derivation of the in-medium nucleon-nucleon cross
section}
The framework for deriving the N-N cross section in
the medium at finite temperature
will be the Matsubara-Green's function
technique \cite{Fetter} (see also \cite{GB})
with non-relativistic propagators.
The two-nucleon scattering in the medium is described by the
thermodynamic T-matrix, which is governed by the corresponding
Bethe-Salpeter equation
\be\label{2.27}
T(121\pr2\pr)=K(121\pr2\pr)+
\int d3\ d3\pr\ d4\ d4\pr\ K(1234)\ G_1(33\pr)\ G_1(44\pr)\
T(3\pr4\pr1\pr2\pr),
\ee
$1$ denotes wavenumber $k_1$, spin $\sigma_1$ and isospin
$\tau_1$ of the nucleons.
The ladder approximation for the thermodynamic T-matrix is
obtained replacing the four-point interaction $K$ in
(\ref{2.27}) by the bare nucleon-nucleon-interaction $V$.
Within the quasiparticle approximation \cite{Fetter}
the product of the two one-particle Greens functions $G_1$ in
(\ref{2.27}) yields
\be\label{3.3}
G^0_2(k_1,k_2,z)
      =\frac{1-f(\epsilon(k_1))-f(\epsilon(k_2))}{z-
      \epsilon(k_1)-\epsilon(k_2)}
\ee
where $z$ is the two-particle energy,
$f(\epsilon)$ is the Fermi distribution function and
$\epsilon(k_1), \epsilon(k_2)$
are the quasiparticle energies.
They are defined in terms of the self energy as
\bea
\epsilon(k_1)&=&\frac{\hbar^2 k_1^2}{2 m}+v(k_1)\nonumber\\
\eea
with
\bea\label{3.2}
v(k_1)&=&
{\rm Re} \Sigma(k_1,\omega)
\mid_{\omega=\epsilon(k_1)},
\eea
where the self energy $\Sigma(k_1,\omega)$ was calculated
in the T-matrix approximation (for details see
ref.\cite{Schmidt}).
%%%%%%%%%%%%%%%%%%%%%%%%%%%%%%%%%%%%%%%%%%%%%%%%%%%%%%%%%%%%%%%%%%%%

Within these approximations two medium effects are contained in
the quantity $G_2^0$.
One of these effects is the phase space occupation $Q$
(Pauli blocking) of the
surrounding nucleons given by the
$Q(k_1,k_2)=1-f(\epsilon(k_1))-f(\epsilon(k_2))$ in eq. (\ref{3.3}).
This form of the Pauli operator takes hole-hole-scattering into
account, that is neglected in the usual Brueckner theory
taking the Pauli operator as
$Q_B(k_1,k_2)=(1-f(\epsilon(k_1)))
(1-f(\epsilon(k_2)))$
\cite{Fetter}.
The possibility for the Pauli operator to
change its sign turns out to be crucial for the onset of
superfluidity (see discussion below).
The second medium contribution is due to the renormalization of
the quasiparticle energies (\ref{3.2})
entering eq. (\ref{3.3}).
%%%%%%%%%%%%%%%%%%%%%%%%%%%%%%%%%%%%%%%%%%%%%%%%%%%%%%%%%%%%%%%%%%%%%%
In the calculation of the in-medium cross section
we used the Paris potential as the bare nucleon-nucleon
interaction.
The Paris potential was
derived from meson theory and gives a quantitatively reliable
description of the on- and off-shell properties of the
nucleon-nucleon interaction in the vacuum \cite{Lacom}. In particular, the
nucleon-nucleon scattering phase shifts which are well known from
experiment are reproduced with a high accuracy.
It has been applied to the calculation of the equilibrium
properties of nuclear matter as well (see ref. \cite{Bal}).

After a partial wave decomposition the thermodynamic T-matrix in the
channel $\alpha=(S,L,J)$ reads in Matsubara-Fourier
representation
\be\label{3.9}
T^{L L'}_{\alpha}(k,k',K,z)=V^{L L'}_{\alpha}(k,k')+\sum_{k'' L''}
V^{L L''}_{\alpha}(k,k'')
G^0_{2}(k'',K,z) T^{L'' L'}_{\alpha}(k'',k',K,z).
\ee
Relative and center-of-mass coordinates were
introduced in eq. (\ref{3.9}) according to $\vec K=\vec k_1+\vec k_2$ and
$\vec k = (\vec k_1-\vec k_2)/2$.
In eq.
(\ref{3.3}) the usual angle averaging
of the Pauli operator $Q(\vec k, \vec K)$ and the quasiparticle
energies $\epsilon(\vec k, \vec K)$ had been carried out.
Furthermore an k-dependent effective two-particle mass
$m_{12}^*(k,K)$ was introduced for the evaluation of the
quasiparticle energies (see \cite{Schmidt} for details).
Thus, it is possible to define an effective chemical potential
$\mu_{rel}=\mu-\Delta \epsilon$
relative to the continuum edge, that incorporates the averaged
single-particle self energy shift
$\Delta \epsilon$ \cite{Alm1}.
Having the T-matrix (\ref{3.9}) at our disposal, generalized
scattering phase shifts may be defined in the following way
(for uncoupled channels)
\cite{Schmidt}
\be\label{3.20}
\pi N(E,K,\mu,T) Q(k,K) T_{\alpha}(k,k,K,E,\mu,T)=
\sin \delta_{\alpha}(E,K,\mu,T)
e^{i \delta_{\alpha}(E,K,\mu,T)},
\ee
with the generalized density of states
\be\label{N}
N(E,K,\mu,T)=\frac{k m^{\ast}_{12}(k,K) }
{\hbar^2 2(2\pi)^3},
\ee
and the relative energy
\be
E=\frac{\hbar^2 k^2}{m^*_{12}(k,K)}.
\ee
The in-medium scattering phase shifts depend on the temperature $T$ and
the  chemical
potential $\mu$ of the medium as well as on the total momentum
$K$ of the pair of nucleons.
In the low-density limit $\mu/T \longrightarrow
-\infty$ the thermodynamic T-matrix (\ref{3.9}) approaches the scattering
T-matrix describing the isolated elastic N-N-scattering.
Correspondingly, the in-medium scattering phase shifts (\ref{3.20}) in this
limit approach the free N-N-scattering phase shifts.

Using the partial wave decomposition, which becomes possible
after angle averaging of the Pauli operator and the
quasiparticle energies, the total thermodynamic
T-matrix may be constructed from the partial T-matrices.
Suppressing isospin indices and
the parametric dependence on $\mu$, $T$ and $K$ the
partial wave decomposition reads
\bea\label{3.8}
T(\vec k, S, M_S, \vec k', S', M_S',z)&=&
\sum_{J,L,L\pr,M_LM_{L\pr}M_J}
T^{LL'}_{\alpha}(k,k\pr,z)\\ \nonumber
& &\times (SLM_SM_L \mid JM_J) Y^{M_L}_L(\hat{k})
(S\pr L\pr M_{S\pr }M_{L\pr } \mid J M_J)
Y^{\ast M_{L\pr }}_{L\pr }(\hat{k\pr }).
\eea
Using eq. (\ref{3.8}) the in-medium differential cross section
for an unpolarized system is defined via the on-shell T-matrix
($|\vec k|=|\vec k'|=k$) as
\be\label{sigpar}
\frac{d \sigma}{d \Omega}(k)=
\frac{N(k)^2}{(2s_1+1)(2s_2+1)}
\sum_{S,M_S, S',M_S'}
\frac{(2 \pi)^4}{k^2}
|T(\vec k S M_S, \vec k' S' M_S')|^2.
\ee
Integrating eq. (\ref{sigpar}) over the angle, one arrives at the total
cross section in the medium
\be\label{sigto}
\sigma(k)=
\sum_{J,L,L'}
\frac{(2J+1) 2\pi^3 N(k)^2 }{(2s_1+1)(2s_2+1) k^2}
|T_{\alpha}^{L L'}(k,k)|^2.
\ee
Eq. (\ref{sigto}) gives the N-N cross section with Pauli blocking in
the intermediate states only; i.e. without correction for Pauli blocking
in the outgoing channel \cite{Mal}.
%%%%%%%%%%%%%%%%%%%%%%%%%%%%%%%%%%%%%%%%%%%%%%%%%%%%%%%%%%%%%%%%%%%%%%%%%%%

For the numerical evaluation of $T^{LL'}_{\alpha}$ and the cross
section (\ref{sigpar}, \ref{sigto})
 we use a separable approximation
of the Paris nucleon-nucleon potential \cite{Plessas}.  The
features of the Paris interaction mentioned above, in particular
the reliable description of the empirical two-nucleon scattering
data,
are preserved in the separable approximation by Plessas
et al.\cite{Plessas}. This separable approximation
was applied in nuclear matter calculations
too \cite{Schmidt,Baldo}.
The general form of a separable interaction is given by
\be\label{3.10}
V^{LL'}_{\alpha}(k,k')=\sum^{N}_{i,j=1} v^{L}_{\alpha i}(k)
\lambda_{\alpha ij} v^{L'}_{\alpha j}(k'),
\ee
For the detailed form of the form factors $v^{L}_{\alpha i}(k)$ and
the coupling strength $\lambda_{\alpha ij}$
for the
separable representation of the Paris
potential see ref. \cite{Plessas}.
The ansatz (\ref{3.10})  permits an analytic solution of the T-matrix
equation \cite{Schmidt} which reads
\be\label{3.13}
T^{L L'}_{\alpha}(kk'K,z)=\sum^N_{ijn} v^L_{\alpha i}(k)
[1-J_{\alpha}(K,\mu,T,z)]^{-1}_{in} \lambda_{\alpha nj} v^{L'}_{\alpha
j}(k'),
\end{equation}
with
\begin{equation}\label{J}
J_{\alpha}(K,\mu,T,z)_{ij}=-4 \pi \int \frac{dk k^2}{(2\pi)^3}
\sum_{n L} \lambda_{\alpha in} v^{L}_{\alpha n}(k)
v^{L}_{\alpha j}(k) G_2^0(k,K,z).
\ee
The total cross section
is calculated introducing (\ref{3.13}) into eq.
(\ref{sigto}).
%%%%%%%%%%%%%%%%%%%%%%%%%%%%%%%%%%%%%%%%%%%%%%%%%%%%%%%%%%%%%%%%%%%%%%%%%
\section{Evaluation of the in-medium N-N cross section}
%%%%%%%%%%%%%%%%%%%%%%%%%%%%%%%%%%%%%%%%%%%%%%%%%%%%%%%%%%%%%%%%%%%%%%%%%
We will give results for the in-medium
total N-N cross section
which is expressed by the
partial T-matrices (\ref{3.13}).

The in-medium total nucleon-nucleon
cross section (\ref{sigto}) depends
on the relative energy $E$ (or the energy
in the laboratory-frame $E_{{\rm LAB}}=2 E$, where one
nucleon is at rest), on the
thermodynamic parameters characterizing the medium, namely the
density $n$ (in units of the saturation density
$n_0=0.17 $fm$^{-3}$),
the temperature $T$ and via the Pauli blocking
$Q(k,K)$ and the selfenergy shifts on the
total momentum $K$ of the pair.
The dependence of the in-medium cross section
$\sigma=\sigma_{np}+\sigma_{nn}$
on the different parameters is illustrated in the following 4
figures.
For comparison the free total N-N cross section, as calculated from
(\ref{sigto}) neglecting all medium effects, was also plotted in
Figs. 1-4 (solid line). Taking into account partial waves up to
$L=2$ we were able to reproduce the experimental free N-N cross
section.

In Fig. 1 we present the total nucleon-nucleon cross section
as a function of the laboratory energy $E_{{\rm LAB}}$ for
various density values at a temperature $T=10$ MeV and a total
momentum of the pair $K=0$.
At a density $n=0.1 n_0$ the in-medium cross section is still
close to the free one showing a slight enhancement.
At $n=0.5 n_0$ we observe a strong
suppression of the cross section at very low energies.
This is mainly due to the Pauli blocking reducing the
available phase space for scattering.
In the
energy range $\sim 50$MeV$<E_{{\rm LAB}}< \sim 150$ MeV a characteristic
enhancement with a maximum at $E_{{\rm LAB}}=80$ MeV
occurs.
This enhancement will be discussed in detail in the next section.
At energies
$E_{{\rm LAB}}> \sim 150$ MeV the in-medium cross section is
slightly suppressed
compared to the free one.
The qualitative behaviour of the in-medium cross section for
$n=n_0$ is the same as for $n=0.5 n_0$. However, the maximum of
the cross section is shifted to higher energies.

%%%%%%%%%%%%%%%%%%%%%%%%%%%%%%%%%%%%%%%%%%%%%%%%%%%%%%%%%%%%%%%%%%%%%%%
In order to study the temperature dependence of the enhancement
of the in-medium cross section as given in Fig.1,
we plotted in Fig. 2 the total in-medium cross section as a
function of $E_{{\rm LAB}}$ for various temperatures at a fixed density
$n=0.5 n_0$ for pairs with total momentum $K=0$.
For temperatures higher than $T=20$ MeV the
cross section approaches the free one and for
$T=50$ MeV (not indicated)
 there is only a very slight deviation from the
free cross section.

With decreasing temperature the cross section is strongly
modified.
For temperatures below 20 MeV one observes the following
behaviour as a function of the energy:\\
The cross section is suppressed compared to the free one for
low energies $0< E_{{\rm LAB}}< \sim 50$ MeV and for high energies
$E_{{\rm LAB}}> \sim 130$ MeV.
Within the energy range $\sim 50<E_{{\rm LAB}}< \sim 130$ the cross section is
enhanced compared to the free one.
This enhancement is especially pronounced for low temperatures.
For $T<10$ MeV a sharp resonance structure develops at
$E_{{\rm LAB}}=90$ MeV (please note that the data in Fig. 2 are plotted
on a logarithmic scale).
For a critical temperature $T_c=4.5$ MeV the in-medium cross section diverges
at this particular energy (see discussion in the next section).

%%%%%%%%%%%%%%%%%%%%%%%%%%%%%%%%%%%%%%%%%%%%%%%%%%%%%%%%%%%%%%%%%%%%%%%
In order to demonstrate how this behaviour  is
changed with the density the in medium cross section is shown in Fig. 3
as in Fig.
2 but for a different density $n=0.2 n_0$.
The qualitative behaviour of the cross section is the same as in
Fig. 2. However, the deviations from the free cross section at a
given temperature are smaller for this smaller density value.
Again a resonance structure is observed, the maximum of which
lies at $E_{{\rm LAB}}=46$ MeV.
Below a critical temperature $T_c=4.2$ MeV
again a divergence of the in-medium cross section shows up at this energy
value.
%%%%%%%%%%%%%%%%%%%%%%%%%%%%%%%%%%%%%%%%%%%%%%%%%%%%%%%%%%%%%%%%%%%%%%%%

The in-medium cross section depends on the total momentum $K$ of the
pair via the Pauli blocking.
In order to demonstrate how this dependence changes the results
given in Figs. 1-3 for $K=0$,
we plotted in Fig. 4
the cross section as a function of $E_{{\rm LAB}}$ for several values
of the total momentum at a fixed density $n=0.5 n_0$ and
temperature $T=5$ MeV.
For $K=0$ one finds the pronounced resonance
( see Fig. 2) discussed above.
With increasing total momentum this resonance is broadened until
it disappears for sufficiently high total momenta.
For $K=400$ MeV/c the in-medium cross section approaches the free
cross section.
This behaviour results from the fact that at the corresponding
higher total momenta $K$ of the pair the Pauli blocking $Q(k,K)$ and the
medium contributions to the one-particle energies
are reduced. Thus, at sufficiently high total
momentum $K$ the medium effects are negligible and the in-medium
cross section approaches the free cross section.
%%%%%%%%%%%%%%%%%%%%%%%%%%%%%%%%%%%%%%%%%%%%%%%%%%%%%%%%%%%%%%%%%
\section{The origin of the enhancement in the in-medium cross section
at low temperatures}

For high temperatures (comparable to the  the Fermi energies)
the in-medium cross
section approaches the free one (see e.g. the $T=20$ MeV curve in
Fig.2).
This is due to the fact that the
medium effects (Pauli blocking, selfenergy shifts) are considerably
reduced at
these temperatures.

The modifications of the in-medium cross section are much more
pronounced at low temperatures.
The most prominent effect found in our calculations is a peak
structure dominating the in-medium cross section at low
temperatures.
Finally, we found a critical temperature $T_c$ at which the
in-medium cross section at a given density shows a divergency
at a particular energy for pairs with zero total momentum $K=0$.

This behaviour may be traced back
to the pole structure of the thermodynamic T-matrix (\ref{3.13}).
Thus, one has to investigate the different channels $\alpha$ of the
thermodynamic T-matrix for the possible
occurrence of poles in dependence on temperature and density.
With decreasing temperature such
a pole first occurs at the particular energy $E=2 \mu_{rel}$
($\mu_{rel}=\mu-\Delta \epsilon$,
relative to the continuum edge)
and at the
critical temperature $T_c$ in the $^3S_1-^3D_1$ channel.

In order to investigate, how the divergence of the in-medium cross
section at $T_c$ and the resonance like structure in the cross
sections at temperatures just above $T_c$ are related to this
pole in the S-D-channel we will use the separable
approximation for the T-matrix (\ref{3.13}).
In ref. \cite{Schmidt} an optical theorem for
the thermodynamical T-matrix was derived, which
is given in momentum representation using eq.
(\ref{N}) as
\be\label{ot}
{\rm Im} T_{\alpha}(k,k',E)=
T_{\alpha}(k,k'',E) \pi Q(k'',K) N(E,K)
T^*_{\alpha}(k'',k',E).
\ee
With the help of this generalized optical theorem
(\ref{ot}) for the
thermodynamic T-matrix
the in-medium cross section can be related
to the imaginary part of the thermodynamic T-matrix.
For a given partial wave $\alpha$ (in particular
$\alpha=^3S_1-^3D_1$) the corresponding partial
cross section is given from (\ref{sigto}) as
\be\label{sr1}
\sigma_{\alpha}(k) \sim |T_{\alpha}(k,k)|^2
=Q^{-1}(k,K) {\rm Im} T_{\alpha}(k,k),
\ee
(the indices $L,L'$ as well as the thermodynamic
parameters are omitted),
where for a rank one (N=1 in eq. (\ref{3.13})) separable interaction
\be\label{imt}
{\rm Im} T_{\alpha}(k,k)=
\frac{\lambda_{\alpha} v^2_{\alpha}(k) {\rm Im}
J_{\alpha}(K,\mu,T,E)}
{(1-{\rm Re}J_{\alpha})^2+({\rm Im} J_{\alpha})^2}
\ee
holds.
Evaluating the imaginary part of $J_{\alpha}$ one finds
\bea\label{imj}
{\rm Im} J_{\alpha}(K,\mu,T,E)&=&
-g^{-1}(E+K^2+2 \Delta \epsilon)
\int \frac{d^3k'}{(2 \pi)^3}
f(\epsilon(K/2+k'))f(\epsilon(K/2-k'))\nonumber\\
\times & &\lambda_{\alpha} v^2_{\alpha}(k')
\pi \frac{\delta(k'-k)}{2 \hbar^2 k'/m^*_{12}(k,K)},
\eea
with
\be
E=\frac{\hbar^2 k^2}{m_{12}^*(k,K)}
\ee
using the property
$1-f(\epsilon(k_1))-f(\epsilon(k_2))
=g^{-1}(\epsilon(k_1)+\epsilon(k_2))f(\epsilon(k_1))f(\epsilon(k_2))$
of the Pauli operator. The quantity $g(E)$ is the Bose
distribution function of the two-particle states.
Consequently, at energies $E=2 \mu_{rel}$ ($K=0$) the quantities
${\rm Im} J_{\alpha}$ and ${\rm Im} T_{\alpha}$ (\ref{imt}) vanish.
This zero of ${\rm Im} T_{\alpha}$ is not restricted to the
quasiparticle approximation (\ref{3.3})
, which can easily be demonstrated by
introducing the spectral representation for the full one-particle propagators
\cite{Henning}.
In the numerator of eq. (\ref{sr1}) this zero is compensated by a
corresponding zero from the inverse Pauli operator $Q^{-1}$ in (\ref{sr1}).
However, the second term in the denominator of eq. (\ref{imt}) is
equal to zero at this particular energy.
Consequently, the magnitude of (\ref{sr1}) at $E=2 \mu_{rel}$ is determined by
the
term $(1-{\rm Re} J_{\alpha}(K=0,\mu,T,E=2 \mu_{rel}))^2$ in the denominator of
(\ref{imt}).
A pole of the T-matrix $T_{\alpha}$ ((\ref{3.13}) for a rank one separable
ansatz, $N=1$)
occurs where both terms
${\rm Im}J_{\alpha}$ and $(1-{\rm Re}J_{\alpha})$ are equal to zero.
The second condition
\be\label{Thou}
1-{\rm Re} J_{\alpha}(K=0,\mu,T=T_c,E=2 \mu_{rel})=0
\ee
is fulfilled at the critical temperature $T_c$.
Thus, the closer the temperature gets to the critical
temperature $T_c$ the larger the partial cross section
(\ref{sr1}) at $E=2 \mu_{rel}$ becomes.
It then gives the dominant
contribution to the total cross section and produces
the resonance like structure shown in figs. 2-4.
Finally, at $T=T_c$ the entire denominator of eq. (\ref{imt})
vanishes at $E=2 \mu_{rel}$ and the corresponding partial cross section
diverges.

On the other hand eq. (\ref{Thou}) is identical
with the Thouless criterion
\cite{Thouless}.
It states, that the sum of the
ladder diagrams (\ref{3.9}) converges only above
the critical temperature $T_c$ (\ref{Thou}), which is equivalent with
the critical temperature found from BCS theory \cite{Thouless}.
Consequently, the critical temperature at which the in-medium
cross section for pairs with zero total momentum diverges,
coincides with the critical temperature for the onset of superfluidity.
In particular,
the same condition (\ref{Thou}) holds for the onset of superfluidity
in symmetric nuclear matter \cite{Alm1} and for the divergence of the
in-medium N-N cross section.
This result does not depend on the particular choice of a
separable interaction \cite{Thouless}.
Using a generalized BCS theory for superfluid nuclear matter
including pairing in the S-D channel Baldo et al.
\cite{BaldoSD} found a critical temperature $T_c$ for the
onset of superfluidity in nuclear matter in agreement with
the calculations of ref. \cite{Alm2} using the Thouless criterion (\ref{Thou}).
Thus, the sharp resonance like structure in Figs. 2-4 can be
interpreted as a precursor for superfluidity.

The position of the resonance or the pole respectively is given
according to the Thouless criterion
by $E=2 \mu_{rel}$ ($E_{{\rm LAB}}=2 E$), where the effective chemical
potential $\mu_{rel}$ (including  the quasiparticle shift)
 is related to the
density $n$ by the equation of state
(see \cite{Alm1} for details).
The values of twice the effective
chemical potential $2 \mu_{rel}$, corresponding to the
densities $n=0.5 n_0$ (Fig.2) and $n=0.2 n_0$ (Fig.3),
coincide with the peak positions of the resonances as indicated
in Figs. 2 and 3.

A relation between two-particle fluctuations above the critical
temperature for the onset of superconductivity was discussed by
Rickayzen \cite{Rickayzen}.
He found, that the instability of the normal state approaching
$T_c$ from above is signalled by the growth of pair fluctuations
$<c_{-k}c_{k}c_{k}^+c_{-k}^+>$ which tend to infinity at $T_c$.
Considering the general relation between the two-particle
correlation function and the T-matrix it can be shown that
these fluctuations are related to the resonance like behaviour
discussed above for the in-medium cross section.

The occurrence of a singularity at the transition to a superfluid
state has already been obtained by Sj\"oberg \cite{Sjoe} in
the calculation of the quasiparticle interaction in nuclear
matter.

More recently, the divergence of the pair susceptibility at the
critical temperature has been discussed by Schmitt-Rink et al.
\cite{SR} for a two-dimensional Fermi gas.

The results obtained for the in-medium N-N cross section can be
compared with other calculations.
Faessler et al. \cite{Faessler}, \cite{Faessler1} calculated the
in-medium neutron-neutron cross section at zero temperature for
two colliding nuclear matters from the G-matrix.
Although these authors negelected hole-hole scattering and
modified the Pauli operator for two colliding Fermi spheres, they
obtained a similiar resonance like behaviour as shown in Figs.
2-4.  The energies of the peaks in refs.
\cite{Faessler}, \cite{Faessler1} agree with the peak energies,
indicated as $4 \mu_{rel}$ in Figs. 2 and 3, for the same
densities.
This corresponds
to the findings of Vonderfecht et al. \cite{Dickhoff} who
could show that even neglecting hole-hole scattering in the
Pauli-operator a bound pair state in the medium
is found in some density range.
Within our finite-temperature approach the
enhancement of the in-medium cross sections can be related
to the onset of a superfluid phase (see discussion above).

Relativistic calculations of the in-medium N-N cross section
performed by ter Haar et al. \cite{Mal} and Li et al.
\cite{Li} do not show the enhancement of the cross section
discussed above.
The discrepancy to the non-relativistic calculations of the
in-medium N-N cross section in ref. \cite{Faessler} and in
this paper as well as to the related mean free path
\cite{Cugnon}, \cite{Schmidt} may be due to the different model
characteristics.

%%%%%%%%%%%%%%%%%%%%%%%%%%%%%%%%%%%%%%%%%%%%%%%%%%%%%%%%%%%%%%%%%%%%%%%

In conclusion, we would like to summarize our results:
\begin{itemize}
\item[(i)] Whereas for high temperatures (comparable to the Fermi energies)
the in-medium
cross section approaches the free one,
strong deviations from the free cross section were found at
high densities and low temperatures and low total momentum. In
particular, we found that for a given density there is a strong
enhancement of the cross section at low temperature near energies
$E=2 \mu_{rel}$.
\item[(ii)] Approaching a critical temperature $T_c$ from above this
enhancement leads to a
resonance like  structure. At a temperature $T=T_c$ the in-medium
N-N cross section for pairs in the medium with zero total momentum
diverges at the particular energy $E=2 \mu_{rel}$.
\item[(iii)] Using the Thouless criterion \cite{Thouless} we could show
that this divergence happens at the same critical temperature
as the onset of a superfluid phase of nuclear matter.
Consequently, the resonance structure could be interpreted as a
precursor of superfluidity.
\item[(iv)] The strong dependence on the density and the temperature of
the surrounding medium is considerably reduced for pairs with
non-zero total momentum $K$. For high enough total momentum the
in-medium cross section approaches the free one.
\end{itemize}

It should be mentioned in which way the approximations we used
could be generalized:

For the evaluation of the T-matrix (\ref{2.27}) the ladder
approximation was used. Thus, the modification of the
N-N-interaction in a dense medium (e.g. higher order
corrections such as screening) was neglected.

Furthermore our calculation is based on the quasiparticle
approximation (\ref{3.3}).
However, the generalization with a finite width
of the one-particle spectral function is in principle straightforward.

It would be interesting to develop a theory which
can describe the in-medium N-N scattering below $T_c$.
This demands approximations going beyond the
quasiparticle approximation.

The determination of the in-medium N-N cross section at
densities above the saturation densities and at higher
energies would require a relativistic treatment such
as the Dirac-Brueckner approach as developed
in refs. \cite{Mal} \cite{Li}.

However, within the standard approximations also used in this
paper, the background for the occurrence of an enhanced
in-medium N-N cross section at relative energies equal to twice
the Fermi energy and temperatures below $20$ MeV can be
understood as a precursor for the onset of superfluidity.
These significant modifications of the in-medium cross section
have to be taken into account in simulations of hot expanding
nuclear matter as produced in heavy-ion reactions.\\
\newpage

{\bf Acknowledgement}\\
We thank W. Bauer, B.L. Friman, P.A. Henning and A.
Sedrakian
for constructive discussions and valuable comments.
Th. A. and G.R. thank the members of the
theory groups at the GSI Darmstadt and at the MSU-NSCL East Lansing  for the
kind hospitality extended to them during their stays.
%%%%%%%%%%%%%%%%%%%%%%%%%%%%%%%%%%%%%%%%%%%%%%%%%%%%%%%%%%%%%%%%%%%%%%%%%%

\newpage
\begin{center}
{\bf \large Figure Captions}
\end{center}
Fig. 1:\\
The in-medium total nucleon-nucleon cross section
$\sigma$ as a function
of $E_{{\rm LAB}}$ at given temperature $T=10$ MeV and total
momentum $K=0$
for several values of the density $n$ (in units of the saturation
density $n_0=0.17$ fm$^{-3}$).
The solid line gives the total free cross section.

Fig. 2:\\
The in-medium total nucleon-nucleon cross section $\sigma$
as a function
of $E_{{\rm LAB}}$ at a given density $n=0.5 n_0$ and total
momentum $K=0$
for several values of the temperature $T$.
The solid line gives the total free cross section.
$4 \mu_{rel}$ denotes the position of the effective chemical
potential as defined in the text.
The critical temperature at which the in-medium cross section
diverges is $T_c=4.5$ MeV.

Fig. 3:\\
The same as in Fig. 2 for a density $n=0.2 n_0$.
The critical temperature at which the in-medium cross section
diverges is $T_c=4.2$ MeV.

Fig. 4:\\
The in-medium total nucleon-nucleon cross section $\sigma$ as a function
of $E_{{\rm LAB}}$ at given density $n=0.5 n_0$ and
temperature $T=5$ MeV
for several values of the total momentum $K$.
The solid line gives the total free cross section.
$4 \mu_{rel}$ denotes the position of the effective chemical
potential as defined in the text.
\newpage
{\Large \bf T. Alm et al., Fig. 1}

{\Large \bf T. Alm et al., Fig. 2}

{\Large \bf T. Alm et al., Fig. 3}

{\Large \bf T. Alm et al., Fig. 4}

{\Large \bf Prof. S.M. Austin\\
Editorial Office Physical Review C\\
1 Research Road\\
Box 1000\\
Ridge NY, 11961\\
U.S.A.\\}

{\Large  Dr. T. Alm\\
Arbeitsgruppe der MPG\\
Theoretische Vielteilchenphysik\\
Universit\"at Rostock\\
Universit\"atsplatz 1\\
18051 Rostock\\
Germany}

\begin{references}
\bibitem{Cassing}
W. Cassing, W. Metag, U. Mosel and K. Niita,
Phys. Rep. {\bf 188}, 363 (1990).
\bibitem{Ogilvie}
C.A. Ogilvie, W. Bauer, D.A. Cebra et al., Phys. Rev. {\bf C
42}, R10 (1990).
\bibitem{Xu}
H. M. Xu, Phys. Rev.  {\bf C 46}, R389 (1992).
\bibitem{Klakow}
D. Klakow, G. Welke and W. Bauer,
Phys. Rev. {\bf C 48}, 1982 (1993).
\bibitem {Cugnon}
J. Cugnon, A. Lejeune, and P. Grange, Phys. Rev. {\bf C 35}, 861
(1987).
\bibitem{Schmidt}
M. Schmidt, G. R\"opke and H. Schulz, Ann. Phys. (N.Y.) {\bf 202}
, 57 (1990).
\bibitem{Faessler}
A. Faessler, Nucl. Phys. {\bf A 495}, 103c (1989).
\bibitem{Faessler1}
A. Bohnet, N. Ohtsuka, J. Aichelin, R. Linden and A. Faessler,
Nucl. Phys. {\bf A 494}, 349 (1989).
\bibitem{Mal}
B. ter Haar and R. Malfliet,
Phys. Rev. {\bf C 36}, 1611 (1987).
\bibitem{Li}
G.Q. Li and R. Machleidt,
Phys. Rev. {\bf C 48}, 1702 (1993).

\bibitem{AlmB}
T. Alm, G. R\"opke , W. Bauer  and M. Schmidt,
to be published.
\bibitem{Baldo}
M. Baldo, J. Cugnon, A. Lejeune and U. Lombardo,
Nucl. Phys. {\bf A 515}, 409 (1990).
\bibitem{Thouless}
D.J. Thouless , Ann. Phys. (N.Y.) {\bf 10}, 553 (1960).
\bibitem{Alm1}
T. Alm ,B.L. Friman, G. R\"opke , and H. Schulz,
Nucl.Phys. {\bf A 551}, 45 (1993).

\bibitem{Fetter}
A.L. Fetter and J.D. Walecka, Quantum Theory of Many-Particle
Systems, McGraw-Hill (1971).
\bibitem{GB}
W.D. Kraeft, D. Kremp, W. Ebeling and G. R\"opke,
Quantum Statistics of Charged Particle Systems
Plenum N.Y. (1986).

\bibitem{Lacom}
M. Lacombe, B. Loiseau, J.M. Richard, R. Vinh Mau, J. Pires and
R. de Tourreil, Phys. Rev. {\bf C 21}, 861 (1980).
\bibitem{Bal}
M. Baldo, I. Bombaci, G. Giansiracusa, U. Lombardo, C. Mahaux,
and R. Sartor,
Phys. Rev. {\bf C 41}, 1748 (1990).
\bibitem{Plessas}
J. Haidenbauer  and W. Plessas,
Phys. Rev. {\bf C 30}, 1822 (1984).
\bibitem{Henning}
P.A. Henning,
Phys. Lett. {\bf B 313}, 341 (1993)
\bibitem{BaldoSD}
M. Baldo, I. Bombaci and U. Lombardo,
Phys. Lett. {\bf B 283}, 8 (1992).
\bibitem{Alm2}
T. Alm, G. R\"opke and M. Schmidt,
Z. Phys. {\bf A 337}, 355 (1990)
\bibitem{Rickayzen}
G. Rickayzen,
in: Green's Functions and Condensed Matter,
Academic Press (1980).
\bibitem{Sjoe}
O. Sj\"oberg,
Nucl. Phys. {\bf A 209}, 363 (1973).
\bibitem{SR}
S. Schmitt-Rink, C.M. Varma and A.E. Ruckenstein,
Phys. Rev. Lett. {\bf 63}, 445 (1989).
\bibitem{Dickhoff}
B.E. Vonderfecht, C.C. Gearhart and W.H. Dickhoff,
Phys. Lett. {\bf B 253}, 1 (1991).
\end{references}
\end{document}